\documentstyle[aps,epsfig]{revtex}
\begin{document}
\draft


\title{Numerical Evidence for the Haldane Conjecture}
\author{B.~All\'es$^{\rm a}$, A. Papa$^{\rm b}$}

\address{$^{\rm a}$INFN Sezione di Pisa, Pisa, Italy}

\address{ $^{\rm b}$Dipartimento di Fisica, Universit\`a della Calabria and
  INFN Gruppo Collegato di Cosenza, Arcavacata di Rende (Cosenza), Italy}

\maketitle

\begin{abstract}
The Haldane conjecture, when applied to the Heisenberg O(3)
model with a $\theta$ term in two dimensions, states that the
correlation length $\xi$ diverges when $\theta$ approaches $\pi$.
To verify this conjecture we have numerically simulated the model
at imaginary $\theta$ and then analytically continued the results
to real $\theta$. We have obtained that the value where the model
should become critical is $\theta=3.10(5)$ in agreement with the
expectation.
\end{abstract}



\section{Introduction}

It has been shown by Haldane that, depending on the value of the spin $\sigma$, the
corresponding one dimensional antiferromagnetic chains of quantum spins present
two kinds of large distance behavior. If $\sigma$
is half--integer, they have a power law correlation function.
Instead for integer spins they lie in a disordered phase and
present an exponentially falling correlation function~\cite{haldane1,haldane2}.
These results were obtained in the limit of large $\sigma$.

The generalization of the above behavior for all values (large or small)
of the spin $\sigma$ is called Haldane conjecture. This conjecture has been widely
studied. Actually a partial result had already been proved for $\sigma=\frac{1}{2}$
in~\cite{lieb}, while for all half--integer spins it was shown
to be correct in~\cite{affleck1}. Moreover, the analytic proof
for spin $\sigma=1$ was given in~\cite{affleck2}.

On the other hand there are indications~\cite{haldane2,shankar,zamolodchikov}
that the one dimensional
antiferromagnetic chain of quantum spins $\sigma$ shares the same large distance
physical properties of the two dimensional O(3)
nonlinear sigma model for classical spins with a
$\theta$ term for $\theta=2\pi \sigma$.
This equivalence would imply that while the ground state of the two dimensional
O(3) nonlinear sigma model at vanishing $\theta$ must display no
long--range order and only short--range spin correlations,
the model at $\theta=\pi$ should be critical.
The first result is well--known, both analytically~\cite{hasenfratz}
and numerically~\cite{alles1}. However the large distance
behavior of the correlation function of the model in the second case
is a not so clearly settled question.

The two dimensional O(3) nonlinear sigma model is a valuable
representation of several types of physical problems. Apart from the
one dimensional quantum spin chains, in condensed matter
physics it may describe the quantum Hall effect as well as being useful to understand
superconductivity~\cite{fradkin}. In particle physics it has in common with
nonabelian gauge theories some important properties such as instantons, asymptotic
freedom (criticality at zero temperature), a $\theta$ term, spontaneous generation of mass, etc.

Two recent numerical calculations of the partition
function for the O(3) model in the presence of a $\theta$
term~\cite{bietenholz,azcoiti} suggest that the theory
undergoes a second order phase transition at $\theta=\pi$
although the two analyses disagree about the universality class. Indeed
the analysis of Bietenholz et al.~\cite{bietenholz} confirms the critical exponents
of the Wess--Zumino--Novikov--Witten model at topological coupling $k=1$
as predicted by Zamolodchikov et al.~\cite{zamolodchikov},
while the numerical study of
Azcoiti et al.~\cite{azcoiti} yields a set of continuously varying critical exponents.

In this work we introduce a direct numerical method
to verify the Haldane conjecture for the two dimensional O(3) nonlinear
sigma model at nonzero $\theta$. The idea is to perform a Monte Carlo simulation to
calculate the correlation length $\xi$ on the lattice as a function of
the $\theta$ parameter and to show that it diverges
at a precise value of $\theta$, called $\theta_{\rm end}$,
which, following Haldane, should be $\theta_{\rm end}=\pi$.

Due to the (suppossed) divergence of the correlation length
at $\theta_{\rm end}$, a direct simulation would become impracticable
as it would require exponentially large lattice sizes.
Moreover the Boltzmann weight in the partition function becomes
complex for real $\theta$ and consequently it loses its probability meaning,
thus precluding the importance sampling of Monte Carlo methods.
We overcome these two difficulties by simulating the theory
at imaginary $\theta$ (where $\xi$ turns out to be small enough
to allow the use of moderate lattice sizes) and analytically
continuing the results to the real $\theta$ values. To this end
we introduce a new fast cluster algorithm that works for imaginary
nonzero theta.
This work is an extended version of the paper appeared
in~\cite{PRDAllesPapa}.

In the next section we shall discuss the formulation of the model on
the lattice and the corresponding lattice definition of the
$\theta$ term and its meaning. The method to calculate the $\theta$
term is introduced in Section~\ref{thetacalc}. The new cluster algorithm
expressly devised for the present work shall
be described in Section~\ref{cluster}. The results and corresponding
plots are displayed in Section~\ref{results}. We end the paper with some
conclusive comments in Section~\ref{conclusions}.

\section{Lattice implementation and notation}
\label{lattice}

The Boltzmann weight of the partition function in our simulations was
$\;\;\exp\left(-S\right)\;\;$ with
\begin{eqnarray}
 S&=&A - i \theta Q \;, \nonumber \\
 A&\equiv&-\beta \sum_{x,\mu} \vec{\phi}(x) \cdot \vec{\phi}(x+\widehat{\mu}) \;,
\label{latticeaction}
\end{eqnarray}
where the sum is extended over all lattice sites $x$ and directions $\mu=1,2$.
The factor $\beta$ is the inverse temperature (in units of the spin coupling),
$Q$ is the total topological charge
or winding number of the configuration (see later) and
$\vec{\phi}(x)$ is a 3--component unit vector that represents
the dynamical variable, a classical spin, at the site $x$.
We have used a square lattice of lateral size $L$ with periodic
boundary conditions.

In the limit where the lattice spacing $a$ vanishes (keeping $L\cdot a$ fixed),
the above expression
for $A$ becomes the action of the classical field theory defined on a
continuum two dimensional plane
\begin{equation}
A\;\,{\buildrel {a\to 0}\over \longrightarrow}\;\, \frac{\beta}{2}\int
 \left(\partial_\mu\vec{\phi}(x)\right)^2\,{\rm d}^2x\; ,
\label{Acontinuum}
\end{equation}
together with the condition $\vec{\phi}(x)^2=1$.

\begin{figure}[htbp]
\centerline{\epsfig{file=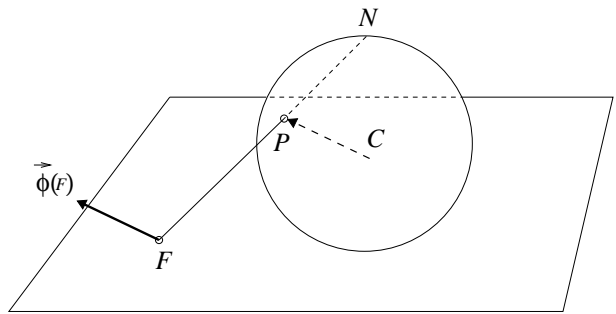,width=0.45\textwidth}}
\vskip 5mm
\caption{A stereographic projection defines an instanton on the
two dimensional O(3) nonlinear sigma model.}
\label{Fig0}
\end{figure}

In Fig.~\ref{Fig0} we show a stereographic projection that defines an instanton
configuration on the O(3) nonlinear sigma model. The two dimensional plane is
shown where the configuration of spins lies.
There is a unit sphere resting on the origin of the plane.
One can draw straight lines that join the north pole $N$ of the sphere with
an arbitrary point $F$ on the plane. One such a line pierces the sphere surface at
point $P$. The unit vector that begins at the center $C$ of the sphere and
points to $P$ is the value of the spin vector $\vec{\phi}(F)$ to be assigned
at the point $F$ of the plane. This construction defines a configuration called
instanton and its analytical expression is the following (the lattice has
been replaced by a continuum by sending $a\to 0$ as above, furthermore
polar coordinates $r$ and $\varphi$ are used to locate the position of the spin
variable)
\begin{equation}
\vec{\phi}(r,\varphi)=\left(\frac{4 r \cos\varphi}{r^2 + 4},\;
\frac{4 r \sin\varphi}{r^2 + 4},\; \frac{r^2 - 4}{r^2 + 4}\right)\;.
\label{instphi}
\end{equation}
In particular, notice that all spins at infinity are identified with
the same value $\vec{\phi}(r=\infty ,\varphi)=(0,0,1)$. Then spins
slowly rotate while approaching $r=0$ until becoming
$\vec{\phi}(r=0 ,\varphi)=(0,0,-1)$ at the origin.

The spin value in Eq.(\ref{instphi}) is a solution of the classical
field equations associated to the action Eq.(\ref{Acontinuum}).

\vskip 1cm

The main feature of instanton configurations is that
the set of all spin vectors describes a complete winding of the sphere, as it
is obvious in the example shown above. This winding can be calculated by
the integral~\cite{rajaraman} (valid in the continuum two dimensional plane)
\begin{eqnarray}
 Q&=&\int \hbox{d}^2x \,Q(x)\;, \nonumber\\
 Q(x)&\equiv&
      \frac{1}{8\pi} \epsilon^{\mu\nu} \epsilon_{bcd}
      \phi^b(x) \partial_\mu \phi^c(x)
      \partial_\nu \phi^d(x)\; ,
\label{continuumaction}
\end{eqnarray}
where $Q$ is called topological charge or winding number and
$Q(x)$ is the topological charge
density. Spatial indices $\mu ,\nu$ and O(3) vector indices $b,c,d$
are summed up. For the instanton of Eq.(\ref{instphi}) it
yields $-1$. There are however many more instantonic configurations,
besides that shown in Fig.~\ref{Fig0} and in general $Q$ takes any positive,
negative or null integer value, depending on how many times and in what
direction the whole set of spins covers the unit sphere.

The second main property is that instanton configurations carry a
finite amount of energy (the operator $A$ in Eq.(\ref{Acontinuum}) takes
on a finite value) even when the lattice size $L$ diverges.

In the Monte Carlo simulations we have used two different definitions of $Q(x)$.
The first one~\cite{papa}
\begin{eqnarray}
Q^{(1)}(x)&\equiv&\frac{1}{32\pi} \epsilon^{\mu\nu} \epsilon_{bcd}\phi^b(x)
       \Big(\phi^c(x + \widehat{\mu}) - \phi^c(x - \widehat{\mu})\Big)
    \cdot \nonumber \\
     &&  \Big(\phi^d(x + \widehat{\nu}) - \phi^d(x - \widehat{\nu})\Big)\; ,
\label{latticeQ1}
\end{eqnarray}
is a symmetrical discretization of the expression 
for $Q(x)$ in Eq.(\ref{continuumaction}) and is usually called
``naive'' definition. The corresponding winding number is
$Q^{(1)}=\sum_x Q^{(1)}(x)$.

The second lattice expression that we used in our simulations
is defined on triangles (not on single sites).
Every plaquette of a square lattice can be
cut through a diagonal into two triangles. If we call $\vec{\phi}_1$,
$\vec{\phi}_2$ and $\vec{\phi}_3$ the fields at the sites of the
three vertices (numbered counterclockwise) of one of these triangles,
then the fraction of spherical angle subtended by these fields is
$Q^{(2)}(\bigtriangleup)$ and it satisfies~\cite{berg}
\begin{eqnarray}
\exp\left(2\pi i  Q^{(2)}(\bigtriangleup)\right) &=&
     \frac{1}{\rho}\Big(
     1+\vec{\phi}_1 \cdot \vec{\phi}_2 +
        \vec{\phi}_2 \cdot \vec{\phi}_3 +
        \vec{\phi}_3 \cdot \vec{\phi}_1 \nonumber \\
    &&  +i \vec{\phi}_1 \cdot \left(\vec{\phi}_2 \times
\vec{\phi_3}\right)\Big)\;,
\label{latticeQ2}
\end{eqnarray}
where $\rho^2\equiv 2 (1 + \vec{\phi}_1 \cdot \vec{\phi}_2)
(1 + \vec{\phi}_2 \cdot \vec{\phi}_3)
(1 + \vec{\phi}_3 \cdot \vec{\phi}_1)$ and
$Q^{(2)}(\bigtriangleup)\in [-\frac{1}{2},+\frac{1}{2}]$.
The above conditions uniquely determine the portion of spherical angle
subtended by $\vec{\phi}_1$, $\vec{\phi}_2$ and $\vec{\phi}_3$ and the
sum of $Q^{(2)}(\bigtriangleup)$ over all triangles yields
the so--called ``geometrical'' topological charge~$Q^{(2)}$.

The two definitions $Q^{(1,2)}$ belong to the same universality
class. In particular they both satisfy the limit
\begin{equation}
Q^{(1,2)}(x) {\buildrel {a\to 0}\over \longrightarrow} a^2 Q(x)\; ,
\end{equation}
where $a$ is the lattice spacing.

\section{Evaluation of Q}
\label{thetacalc}

In general, a definition of $Q$ on the lattice does
not necessarily lead to integer values on a single configuration.
To recover integer results for $Q$ on ensembles
of configurations that contain the same topological charge,
we must renormalize this operator. The lattice and
the continuum topological charges are related by~\cite{pisa1}
\begin{equation}
 Q^{(1,2)}=Z_Q^{(1,2)}Q\;,
\label{qlzq}
\end{equation}
$Z_Q^{(1,2)}$ being the corresponding renormalization constant.
The origin of this constant can be traced back to the presence of
statistical fluctuations in an otherwise smooth instantonic configuration.
In general operators that reveal the topological charge $Q$ of a
configuration give wrong answers due to the disturbance caused by
the presence of fluctuations.

The $Z_Q^{(1,2)}$ function depends only on the temperature $\beta$ and is chosen
in such a way not to depend on $\theta$ since the introduction of this
term does not modify the structure of fluctuations in the model.
Moreover it satisfies $0<Z_Q^{(1,2)}\leq 1$ for
all values of the lattice spacing~\cite{vicari,papa}.

$Z_Q^{(1,2)}$ can be calculated either in perturbation
theory~\cite{pisa1,vicari,papa} or by a nonperturbative numerical
method~\cite{pisa2,farchioni,beccalles}.
We have used the latter. In a nutshell it works in the following way:
one measures $Q^{(1,2)}$ on an instantonic configuration (topological
charge $Q=+1$) after heating it at a temperature $\beta$. From Eq.(\ref{qlzq})
this measurement yields $Z_Q^{(1,2)}$.

{}First of all
an instanton with topological charge +1 is put by hand on the
lattice. We used the solution~\cite{belavin}
\begin{equation}
\frac{\phi^1 + i \phi^3}{1 - \phi^2} =
\frac{x_1 - L/2 - i (x_2 - L/2)}{\lambda\, {\rm e}^{i\pi/4}}\;,
\end{equation}
where $\phi^b$ is the $b$--component of the field at site $x=(x_1, x_2)$,
$L$ is the size of the lattice and $\lambda$ is the size of the instanton. This
solution is the one shown in Eq.(\ref{instphi}) after centering it in
the middle of the lattice $(L/2,L/2)$, dilating it to the size
$\lambda$ (the size in Eq.(\ref{instphi}) is $\lambda=2$)
and performing appropriate rotations both in $x$--space and O(3) space~\cite{rots}.
In order to work with a rather stable instanton (recall that single
instantons on the lattice are only metastable solutions of the
equations of motion on a torus) it is convenient to choose
$\lambda\lesssim 0.15\, L$ and $\lambda\gtrsim 8$~\cite{beccalles}.
We took $\lambda=16$ on a $L=120$ lattice.

Then 100 updating steps are applied (we used the Heat--Bath
algorithm~\cite{creutz} on the conventional O(3) nonlinear sigma model
without a $\theta$
term since the renormalization constant to be used in Eq.(\ref{qlzq}) does not
depend on $\theta$). After every Heat--Bath step the value of $Q^{(1,2)}$
is measured and, in order to monitor the instantonic contents and check that
it is not varied after the updating step, $Q^{(1,2)}$ is measured again
after 6~relaxation hits applied on a separate copy of the running configuration.
The complete history of 100 Heat--Bath updating steps and related measurements
of $Q^{(1,2)}$ is called a {\it trajectory}. In the calculation of $Z_Q^{(1)}$
we used $4\cdot 10^4$ trajectories at $\beta=1.5$ and $1.6$ and $10^4$
trajectories for $\beta=1.7$ and $1.75$. The average of $Q^{(1,2)}$ on
all trajectories, as long as their topological charge remained equal to $+1$,
yielded $Z_Q^{(1,2)}$.

The O(3) nonlinear sigma model develops an infrared divergence in its instanton
size distribution~\cite{spencer,farchioni}.
This divergence facilitates the copious creation of new
instantonic objects at every updating step, thus modifying the total topological
charge of the configuration. For this reason the relaxation test is extremely
important.

As a relaxation method we used the so--called cooling~\cite{teper}. It
sweeps through the whole lattice and modifies one by one
every single spin variable in order to locally minimize the energy of
the relaxed configuration. Actually many variants of the cooling method
exist in the literature but in practice all of them act analogously~\cite{cosmai}.

The above nonperturbative method is summarized by the expression
\begin{equation}
Z_Q^{(1,2)}=\frac{\int_{\rm 1-instanton} {\cal D}{\vec{\phi}}\; Q^{(1,2)}\;
\exp\left(-A\right)}{\int_{\rm 1-instanton} {\cal D}{\vec{\phi}}\;
\exp\left(-A\right)}\;,
\label{heatingmethod}
\end{equation}
where ${\cal D}{\vec{\phi}}$ stands for the measure
\begin{equation}
\prod_x\left(\delta(\vec{\phi}(x)^2- 1)\;\prod_b {\rm d}\phi^b(x) \right)
\end{equation}
and $\int_{\rm 1-instanton}$ means that the integration is extended over all
configurations (fluctuations) that preserve the background of one instanton.
Since the geometrical charge $Q^{(2)}$ is +1 till the background
configuration is one instanton, the expression~(\ref{heatingmethod})
yields $Z_Q^{(2)}=1$ for all $\beta$~\cite{luscher}. This result derives from
the fact that fluctuations, viewed as local large (positive or negative) values
of the spherical angle in some spherical triangles, cancel out when summing up
all individual contributions $Q^{(2)}(\bigtriangleup)$.

\begin{figure}[htbp]
\centerline{\epsfig{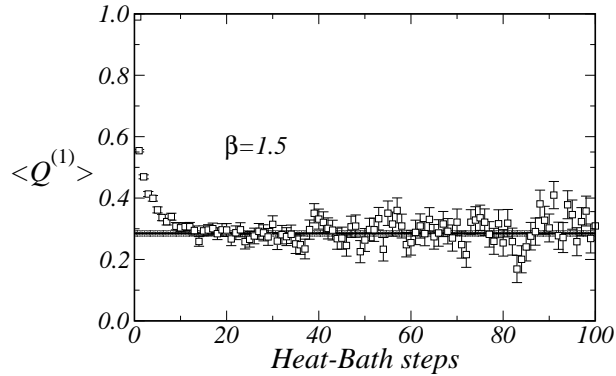}}
\caption{Data for $\langle Q^{(1)}\rangle$ start at $+1$
at the 0--th Heat--Bath step and
then they go down until reaching a plateau. The height of the
horizontal line and grey
band are the value and error respectively of $Z_Q^{(1)}(\beta=1.5)$.}
\label{Fig1}
\end{figure}

The determination of $Z_Q^{(1)}$ is not so trivial and an
example of such an evaluation is shown in Fig.~\ref{Fig1}.
Measures of $Q^{(1)}$ on configurations that have 
topological charge $+1$ attain to a plateau (in general after a few
Heat--Bath steps) and stay on it for the rest of the updating steps.
The height of this plateau is the value of $Z_Q^{(1)}$.
In Table~1 the results for $Z_Q^{(1)}$ at the
values of $\beta$ used in the present work are given.

\vskip 1cm

{{{\rm Table 1.} $Z_Q^{(1)}$ and $\theta_{\rm end}$ for the topological
charge $Q^{(1)}$.}
\vskip 2mm
{\centerline{
\begin{tabular}{|p{10mm}|p{14mm}|p{14mm}|p{18mm}|p{15mm}|p{15mm}|} \hline
$\beta$ & $\left(\overline{\theta}_{{\rm end}}\right)^2$ & $Z_Q^{(1)}$ &
  $\chi^2/$d.o.f. & $\theta_{\rm end}$ \\ \hline
1.5 & 111(5) & 0.285(9) & 0.90 & 3.00(12) \\ \hline
1.6 & 94(5) & 0.325(6)  & 0.45 & 3.15(10) \\ \hline
1.7 & 67(3) & 0.380(6)  & 1.04 & 3.11(9)  \\ \hline
1.75 & 56(3) & 0.412(5) & 0.68 & 3.08(9)  \\ \hline
\end{tabular}
}}}

\vskip 1cm

\vskip 1cm

\begin{figure}[t]
\centerline{\epsfig{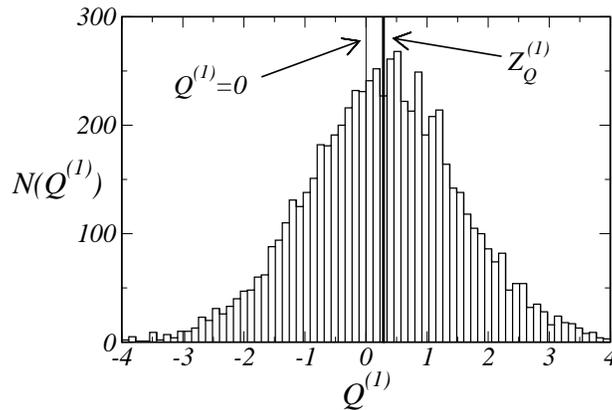}}
\caption{Histogram displaying the distribution of naive topological
charge at $\beta=1.5$ starting from an initial configuration with background
topological charge $+1$. Observe the shift of the gaussian--like
distribution towards positive values of $Q^{(1)}$ (a vertical thin
line indicates the zero value and a thicker line the result of $Z_Q^{(1)}$).}
\label{Fighisto}
\end{figure}

In Fig.~\ref{Fighisto} an histogram of the distribution of
topological charge $Q^{(1)}$ is shown. It has been produced from
the data of $Q^{(1)}$ obtained during the calculation of $Z_Q^{(1)}$
at $\beta=1.5$ and contains 70 bins within the interval
$Q^{(1)}\in\left[-4,+4\right]$. For each trajectory in the calculation
of $Z_Q^{(1)}$ we obtained the average of $Q^{(1)}$ over
the steps that come after the onset of the plateau (from Fig.~\ref{Fig1}
this happens at about the 15th step for $\beta=1.5$) and for which the
cooling test gave a background topological charge $+1$. The histogram
of Fig.~\ref{Fighisto} displays the distribution of these averages.
Each of the above averages turns out to be an uncorrelated
estimate of $Z_Q^{(1)}$. The thick vertical line is the
value of $Z_Q^{(1)}$, 0.285(9). Actually the error in the evaluation
of $Z_Q^{(1)}$ was determined by using this kind of plot. Indeed,
it was extracted by usual gaussian analysis on the histogram.

Recall that the number of trajectories for $\beta=1.5$ was 40000. However
the area of the histogram in Fig.~\ref{Fighisto} is much less than 40000. This is
due to the fact that in many trajectories the background charge is no longer
$+1$ already at the beginning of the plateau. In such cases, the whole
trajectory was discarded. The area under the histogram in Fig.~\ref{Fighisto}
is about 6700. The small ratio 6700/40000 gives an idea of the
frequent creation of new instantonic objects that modify the
background topological charge.

\vskip 1cm

The renormalization of the topological charge brings about a relevant
consequence for our study: the $\theta$ parameter that appears
in the expression of the Hamiltonian used in the computer program
during the simulations in general is {\it not} equal to the
true physical $\theta$ parameter. Henceforth we shall
call $\overline{\theta}$ the parameter that appears in the simulation program
and the relation
among the two parameters is $\theta=\overline{\theta}Z_Q^{(1,2)}$.
The value of $\overline{\theta}$ where the correlation length diverges will
be called $\overline{\theta}_{\rm end}$.
Since $Z_Q^{(2)}=1$, it is clear that this distinction among theta parameters
is irrelevant for the geometrical charge $Q^{(2)}$.

We have simulated the model at several temperatures $\beta$ and
parameters $\overline{\theta}$ for the two topological charge operators $Q^{(1)}$ and
$Q^{(2)}$ in order to show that our results are independent of the operator
chosen for the simulation. Moreover, for the case of the naive topological
charge $Q^{(1)}$, we have introduced a very fast cluster algorithm.
Instead, for the geometrical charge $Q^{(2)}$ a rather slow Metropolis--like
algorithm has been used. Actually the cluster algorithm for the naive charge
was so efficient that it took much less computer time to investigate this charge
than the geometrical one even though the naive charge required the
extra calculation of $Z_Q^{(1)}$ by a separate
out--of--equilibrium simulation for each temperature $\beta$ as described above.

\section{Cluster algorithm for imaginary $\theta$}
\label{cluster}

Although the use of the topological charge density $Q^{(1)}$ requires
the knowledge of a renormalization constant, it brings about the advantage
that the Hamiltonian in~(\ref{latticeaction}) can be simulated on the lattice by
use of a fast cluster algorithm when $\overline{\theta}$ is imaginary.

Let us briefly describe the main characteristics of the new cluster
algorithm expressly devised for the present work.
The first part of an updating step with the usual Wolff algorithm~\cite{wolff}
for the standard O(3) sigma model without a $\theta$ term consists in choosing
a random unit vector $\vec{r}$ in such a way that every dynamical
field can be split in a component parallel to $\vec{r}$ and the rest,
$\vec{\phi}(x)=\left(\vec{\phi}(x)\cdot\vec{r}\right)\vec{r} +
\vec{\phi}_\bot(x)$, where $\vec{\phi}_\bot(x)$ denotes the part of
$\vec{\phi}(x)$ orthogonal to $\vec{r}$. Then the signs of
$\left(\vec{\phi}(x)\cdot\vec{r}\right)$ for all $x$ are updated \`a la
Swendsen--Wang as in the Ising model~\cite{swendsen}.

By introducing the above separation for $\vec{\phi}(x)$ in the
expression~(\ref{latticeQ1}) and recalling elementary properties of
determinants in three dimensional vector spaces, we can rewrite it as
\begin{eqnarray}
 Q^{(1)}(x)&=& \nonumber \\
  \frac{1}{16\pi} &\Big\{&
  \left(\vec{\phi}(x)\cdot\vec{r}\right)
      \big(d_{1,2} + d_{-1,-2} + d_{2,-1} + d_{-2,1}\big) \nonumber \\
  &&+ \left(\vec{\phi}(x+\widehat{1}\,)\cdot\vec{r}\right)
      \left(d_{0,-2} - d_{0,2}\right) \nonumber \\
  &&+\left(\vec{\phi}(x-\widehat{1}\,)\cdot\vec{r}\right)
      \left(d_{0,2} - d_{0,-2}\right) \nonumber \\
  &&+\left(\vec{\phi}(x+\widehat{2}\,)\cdot\vec{r}\right)
      \left(d_{0,1} - d_{0,-1}\right) \nonumber \\
  &&+\left(\vec{\phi}(x-\widehat{2}\,)\cdot\vec{r}\right)
      \left(d_{0,-1} - d_{0,1}\right) \Big\}\;,
\label{latticeQ1bis}
\end{eqnarray}
where $x\pm\widehat{1}$ means the site at the position one step forward
(backward) in the direction ``1'' starting from site $x$
and the notation $d_{i,j}$ stands for the $3\times 3$ determinant
\begin{equation}
d_{i,j}\equiv\det \left( \begin{array}{ccc}
r^1 & r^2 & r^3 \\
         {\phi}^1(x+\widehat{i}\;) &
                   \quad {\phi}^2(x+\widehat{i}\;) &
                   \quad {\phi}^3(x+\widehat{i}\;)\\
         {\phi}^1(x+\widehat{j}\;) &
                   \quad {\phi}^2(x+\widehat{j}\;) &
                   \quad {\phi}^3(x+\widehat{j}\;)\\
\end{array} \right)\;.
\label{definitiond}
\end{equation}
In this fashion the theory at each updating step looks like an Ising model
in the bosom of an external local magnetic field $h(x)$ because the
expression in Eq.(\ref{latticeQ1bis}) is linear in
$\left(\vec{\phi}\cdot\vec{r}\right)$. The value of this field varies at
each updating and accordingly it must be recalculated after every step.
Recall that all Monte Carlo simulations
have been performed with an imaginary parameter $\overline{\theta}=
+i\overline{\vartheta}$, ($\overline{\vartheta}\in{{\rm I}\kern-.23em{\rm R}}$).
By gathering all contributions
of the type shown in Eq.(\ref{latticeQ1bis}) that contain 
$\left(\vec{\phi}(x)\cdot\vec{r}\right)$ at site $x$
one can readily derive the effective magnetic field at this site,
\begin{eqnarray}
  h(x)&=&-\;\frac{\overline{\vartheta}}{16\pi}\vert\vec{\phi}(x)\cdot\vec{r}\;\vert
\Big(d_{1,2} + d_{-1,-2} + d_{2,-1} + d_{-2,1} \nonumber \\
 &+& \;\;\, d_{-1,-1-2} + d_{-1+2,-1} + d_{1,1+2} + d_{1-2,1}
   \nonumber \\
 &+& \;\;\,d_{2,2-1} + d_{2+1,2} + d_{-2,-2+1} + d_{-2-1,-2}\Big)\;.
\label{fieldmag}
\end{eqnarray}
$d_{i+k,j}$ (and analogous terms in~(\ref{fieldmag}))
are the straightforward generalization of the above
definition~(\ref{definitiond}) when the site is obtained by shifting
two steps from the original position~$x$, the first in the direction
$\;\widehat{i}$ and the second in the direction $\widehat{k}\;$.

Hence the last step in the updating consists in applying to the
above expressions an algorithm valid for the Ising model in presence
of a magnetic field. In the literature there are two such algorithms,
the Lauwers--Rittenberg~\cite{lauwers} and the
Wang~\cite{wang,dimi} methods. After testing their
perfomances and comparing the corresponding decorrelation times
with the usual Metropolis~\cite{metropolis},
Heat--Bath~\cite{creutz} and overHeat--Bath~\cite{petronzio},
we decided on the Wang algorithm. It consists in
placing the magnetic field on an extra, fictitious site
(called ghost site or ghost spin) that couples to every Ising
spin through the value
of $h(x)$. Using this coupling on the same footing
as all other terms in the action, the Fortuin--Kasteleyn
clusters~\cite{fortuin} are
created by using the Hoshen--Kopelman algorithm~\cite{hoshen} and
then updated with the usual $\frac{1}{2}$ probability. The only
distinctive feature of the presence of a ghost spin is that the
cluster that contains it does not flip.

\vskip 1cm
\begin{figure}[htbp]
\centerline{\epsfig{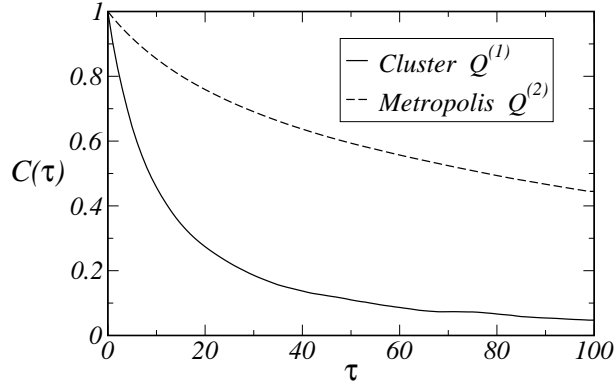}}
\caption{Autocorrelation functions for the Metropolis ($Q^{(2)}$) and
cluster ($Q^{(1)}$) algorithms as a function of the ``updating time''
$\tau$ which has a discrete ticking at each updating step.}
\label{Figcompar}
\end{figure}

Following the proof given in~\cite{wolff}, it can be seen that our
algorithm also satisfies the detailed balance property.

To generate the initial random unit vector $\vec{r}$ the method proposed in
Ref.~\cite{niedermayer} was used.

In Fig.~\ref{Figcompar} we show the autocorrelation functions
\begin{equation}
C(\tau)\equiv\frac{\langle E_0 E_\tau\rangle - \langle E\rangle^2}
{\langle E^2\rangle - \langle E\rangle^2}\;,
\end{equation}
calculated from the measures of the energy operator
$E\equiv\vec{\phi}(x)\cdot\vec{\phi}(x+\widehat{\mu})$ (not summed over $\mu$),
for the two algorithms: Metropolis when the $Q^{(2)}$ operator is used and the
above--described cluster algorithm for $Q^{(1)}$. $E_\tau$ indicates the
$\tau$--th measurement. In both cases $\beta=1.5$ and
$L=120$. The theta parameter was $\overline{\vartheta}=10$ for $Q^{(1)}$
and $\vartheta=2.85$ for $Q^{(2)}$ (note that this choice was
dictated by the condition $\theta=Z_Q^{(1)}(\beta=1.5)\overline{\theta}$).
The plot clearly exhibits the major efficiency of the cluster algorithm.

\section{Results}
\label{results}

As the ground state of the model is a triplet~\cite{controzzi}, we studied the
correlation functions of operators having one O(3) index. We measured the correlation
of the two operators
\begin{equation}
 {\overrightarrow{\cal O}}_1(x)\equiv\vec{\phi}(x)\,,\qquad
 {\overrightarrow{\cal O}}_2(x)\equiv\vec{\phi}(x)\times\vec{\phi}(x+\widehat{1}\,)\,.
\end{equation}
Firstly we calculated the corresponding wall operators by averaging over the
$x_1$ coordinate
\begin{equation}
{\overrightarrow{\cal W}}_1(x_2)\equiv \frac{1}{L}\sum_{x_1}
{\overrightarrow{\cal O}}_1(x) \,,\;
{\overrightarrow{\cal W}}_2(x_2)\equiv \frac{1}{L}\sum_{x_1}
{\overrightarrow{\cal O}}_2(x) \;.
\label{operators}
\end{equation}
In order to extract the correct correlation length and to clean its signal
from any mixture with higher eigenvalues of the Transfer Matrix, we used
the variational method of Ref.~\cite{kronfeld}, where
$\xi$ is obtained from the exponential decay of the
largest eigenvalue of the correlation matrix
\begin{equation}
\langle {\overrightarrow{\cal W}}_i(x_2)\cdot{\overrightarrow{\cal W}}_j(0)\rangle -
\langle{\overrightarrow{\cal W}}_i\rangle \cdot\langle{\overrightarrow{\cal W}}_j\rangle\; .
\label{corrmatrix}
\end{equation}

No improved estimators~\cite{wolff2} were used since the operators in~(\ref{operators})
contain too many fields $\vec{\phi}$ and this fact leads to intractable
sums over clusters~\cite{improestimatorscomment}.

\vskip 1cm

\begin{figure}[htbp]
\centerline{\epsfig{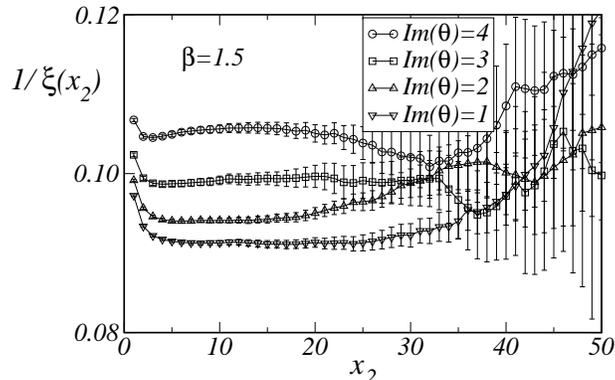}}
\caption{Determination of $\xi$ at $\beta=1.5$ for four values of
$Im(\theta)\equiv\vartheta$ for the case of the $Q^{(1)}$ operator.
Lines are drawn to guide the eye.}
\label{Figmass}
\end{figure}

In Fig.~\ref{Figmass} the method that we followed to extract $\xi$ is
illustrated with four examples. The exponent of the fall--off
of the largest eigenvalue
$\alpha_{\rm max}$ of the correlation matrix, Eq.(\ref{corrmatrix}),
is plotted against the distance $x_2$. For each $x_2$ it was extracted by comparing
the expression
\begin{equation}
\frac{\alpha_{\rm max}(x_2)}{\alpha_{\rm max}(x_2 - 1)}\; ,
\end{equation}
with the theoretical behavior ($L$ is the lattice size)
\begin{equation}
\frac{\cosh\Big(\left(x_2 - L/2\right)/\xi\Big)}
     {\cosh \Big(\left(x_2 - 1 -L/2\right)/\xi\Big)}\; ,
\end{equation}
and its error was determined by jackknife. An approximate plateau
is clearly identified at moderate distances. The definite values of
$\xi$ and its error were chosen self--consistently at $x_2=2\xi$.

{}For each $\beta$ we calculated $\xi$ from Monte Carlo simulations at
several imaginary values of $\overline{\theta}$ for both $Q^{(1)}$ and
$Q^{(2)}$. The set of results for a given $\beta$ were then analytically
continued from imaginary $\overline{\theta}$
to the real $\overline{\theta}$ axis by an usual procedure of numerical
extrapolation. In the extrapolation we avoided using a trial function
dictated by some theoretical argument, like for instance
$1/\xi=c_1 \left(c_2 - \overline{\theta}^2\right)^{2/3}$ which is,
up to logarithmic corrections, the Renormalization Group
prediction~\cite{affleck3}, because
such an analytic form implicitly assumes the vanishing of $1/\xi$ at a precise
value of $\overline{\theta}$ (not to say that it is
supposed to be accurate only in a close neighborhood of its zero,
$\overline{\theta}=\sqrt{c_2}$).
Instead we made the extrapolations by using
polynomials in $\overline{\theta}^2$ and ratios of such polynomials. These
functional forms are indeed both simple and very general and they leave room
for any possible behavior in~$\overline{\theta}$.

\vskip 1cm

\begin{figure}[htbp]
\centerline{\epsfig{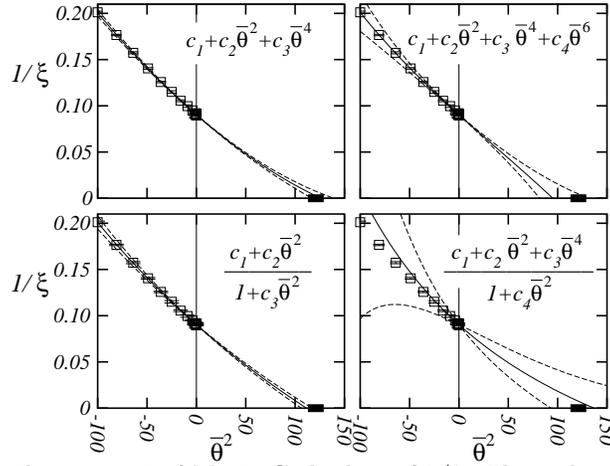}}
\caption{Four extrapolations for the same set of Monte Carlo data of $1/\xi$.
The scale in the axes is the same for the four windows.
Data were extracted from simulations at $\beta=1.5$ and using the $Q^{(1)}$
operator for the topological charge. The horizontal black bar is placed
at the value of $\overline{\theta}$ where the Haldane conjecture predicts
a critical behavior. Each continuous line is the result of
the extrapolation and the dashed lines
enclose the boundary of its error.}
\label{Fig2}
\end{figure}

\vskip 1cm

\subsection{Results for $Q^{(1)}$}

$2\cdot 10^5$ decorrelated propagators were measured for all values of 
$\overline{\theta}$ at each $\beta$. They were obtained after separating
consecutive configurations by a combination of one Heat--Bath, two overHeat--Bath
and one cluster updatings. The values obtained
for $1/\xi$ at $\beta=1.5$ with their error bars are the squares
in Fig.~\ref{Fig2}. In this figure
four different extrapolations are shown (the extrapolation functional forms
are displayed). The Haldane conjecture predicts that $1/\xi$ vanishes at
$\left(\pi/Z_Q^{(1)}(\beta=1.5)\right)^2$. This value is indicated by the
small horizontal shadowed bar on the $\overline{\theta}^2$ axis
(its horizontal width arises from the error in the evaluation
of $Z_Q^{(1)}(\beta=1.5)$, see Table~1). From Fig.~\ref{Fig2} it seems clear that
all analytic continuations are in fair agreement among themselves and with
the Haldane conjecture. We emphasize that no prejudices about the
possible zeroes were included in the extrapolating functions.

\vskip 10mm
\begin{figure}[htbp]
\centerline{\epsfig{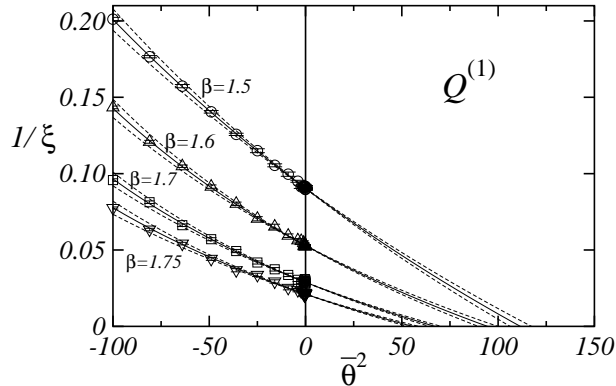}}
\caption{Behavior of $1/\xi$ as a function of $\overline{\theta}^2$.
Circles ($\beta=1.5$), up triangles ($\beta=1.6$), squares ($\beta=1.7$)
and down triangles ($\beta=1.75$) are the data from the simulation at imaginary
$\overline{\theta}$ ($\overline{\theta}^2<0$) by using the $Q^{(1)}$ lattice
topological charge. Meaning of continuous and dashed lines as in Fig.~\ref{Fig2}.}
\label{Fig3}
\end{figure}

{}Four values of $\beta$ were studied in the case of $Q^{(1)}$: $\beta=1.5$,
1.6, 1.7 and 1.75. The respective sets of Monte Carlo data for $1/\xi$ are
shown in Fig.~\ref{Fig3} as circles, up triangles, squares and down
triangles. The lattice sizes were 120, 180, 340 and 470 respectively.
The extrapolations in this figure were done
by using the functional form $(c_1 + c_2\, \overline{\theta}^2)/
(1 + c_3\, \overline{\theta}^2)$ since it was the one
that produced the value of the statistical $\chi^2$ test closer to 1
for all four $\beta$'s
(the values for $\chi^2/$d.o.f. are listed in Table~1).

The results of the analytic continuations are given in Table~1.
The physical value of $\theta$ where the theory becomes critical
is given by $\theta_{\rm end}=\overline{\theta}_{\rm end} Z_Q^{(1)}$.
The numbers in the last column are in good agreement
with the prediction that criticality is achieved when $\theta$ equals $\pi$.
Other functional forms used for the extrapolations led to similar
results although in some cases the $\chi^2$ test was far from
unity and hence the related extrapolation seemed statistically
unlikely (see for instance the $(c_1 + c_2\, \overline{\theta}^2
+ c_3\, \overline{\theta}^4)/(1 + c_4\, \overline{\theta}^2)$
case in Fig.~\ref{Fig2}).

The lattice sizes were chosen large enough to meet at $\overline{\theta}=0$ the
condition $L/\xi\gtrsim 10$ (to be specific, these ratios were 10.8, 9.5,
9.8 and 9.9 for $\beta=1.5$, 1.6, 1.7 and 1.75 respectively).
Once this inequality holds at
$\overline{\theta}=0$, it is amply realized at the values of $\overline{\theta}$ where
the simulations were performed as inferred from Fig.~\ref{Fig3}. This fact
warrants the absence of significant finite size effects.

On the other hand, from the Monte Carlo values of $\xi$ shown in
Fig.~\ref{Fig3} one can see
that we simulated the model at correlation lengths that altogether satisfy
$\xi>5$ which is safely far from the strong coupling region (in this region
the universality property loses its meaning and results may depend on the
choice of operators used in the action).

\subsection{Results for $Q^{(2)}$}

In this case the usual Metropolis algorithm was used for updating.
Observe that when the spin variable $\vec{\phi}(x)$ is updated, the $\theta$
term in Eq.(\ref{latticeaction}) will contribute
to the variation of the Hamiltonian only if the topological charge gets
modified within the only six triangles that surround the site $x$.
Such modifications occur only if some instantonic object rises or
disappears in the area delimited by these six triangles. Such an event
barely occurs on a such a small area and as a consequence prolonged
decorrelations must separate consecutive measurements of any operator
(see Fig.~\ref{Figcompar}).

$10^5$ independent propagators, separated by 100 decorrelation updatings,
were measured for each value of $\theta$
(recall that for $Q^{(2)}$ we have $\overline{\theta}=\theta$). We report
data for only two values of $\beta$. Notice that the total statistics
and the number of $\beta$ values studied is evidently smaller here than in the
previous subsection. As explained above, this is due to the use of a
much less efficient updating algorithm.

\begin{figure}[htbp]
\centerline{\epsfig{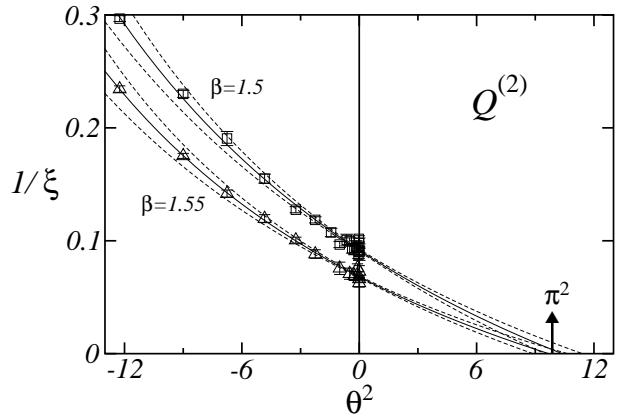}}
\caption{Behavior of $1/\xi$
as a function of $\theta^2$ for the geometrical charge $Q^{(2)}$.
Squares ($\beta=1.5$) and up triangles ($\beta=1.55$).
In this case $\theta=\overline{\theta}$ and the position of $\theta=\pi$ is marked.
Meaning of continuous and dashed lines as in Fig.~\ref{Fig2}.}
\label{Fig4}
\end{figure}

Data are displayed in Fig.~\ref{Fig4} as squares and triangles for $\beta=1.5$
and 1.55 respectively. The corresponding lattice sizes were $L=110$ and 150.
The value of $\theta$ where Haldane predicted
the vanishing of $1/\xi$ is indicated with an arrow, $\theta^2=\pi^2$.
Data near $\theta^2=0$ are very noisy. This is due to the relatively poor
statistics obtained in the simulations with the operator $Q^{(2)}$.

The numerical results are given in Table~2. Comments similar to
the $Q^{(1)}$ case apply to the extrapolations shown in the figure.
{}From Fig.~\ref{Fig4} the correlation lengths satisfy $\xi>3$
which is still away from the strong coupling regime.
Again the results are in fair agreement with the conjecture.

\vskip 1cm

{{{\rm Table 2.} $\theta_{\rm end}$ for the operator $Q^{(2)}$.}
\vskip 2mm
{\centerline{
\begin{tabular}{|p{10mm}|p{16mm}|p{10mm}|p{16mm}|p{16mm}|} \hline
$\beta$ & $\left(\overline{\theta}_{\rm end}\right)^2$ & $Z_Q^{(2)}$ &
  $\chi^2/$d.o.f.  & $\theta_{\rm end}$  \\ \hline
1.5 & 10.4(1.0) & 1.0 & 1.72  & 3.22(16) \\ \hline
1.55 & 9.7(1.0) & 1.0 & 0.73  & 3.11(16) \\ \hline
\end{tabular}
}}}

\vskip 5mm

By averaging all results for both topological charge operators and assuming
gaussian errors we obtain that the model should become critical
at $\theta_{\rm end}=3.10(5)$. This is the chief result of our work.

\vskip 10mm

\section{Conclusions}
\label{conclusions}

We have simulated the O(3) nonlinear sigma model
in two dimensions with an imaginary $\theta$ term at several values of
the inverse temperature $\beta$. The correlation length was measured and extrapolated
towards real $\theta$. In all cases the extrapolation showed a divergence at a
value of $\theta$ compatible with the Haldane conjecture $\theta=\pi$.
Our result is $\theta=3.10(5)$ which agrees within errors with the
conjecture. This value seems very robust as it is
independent of the topological charge density operator chosen for
the simulation. In particular, an operator $Q^{(1)}$ that requires
a nontrivial
renormalization constant leads to the same conclusion than another
operator (the geometrical charge $Q^{(2)}$) that does not renormalize.

A direct numerical study of the model at $\theta$ values which are both
real and close to $\pi$ is unfeasible. First of all at real $\theta$
the Hamiltonian becomes complex, thus preventing the importance sampling
in Monte Carlo methods. Furthermore exponentially large lattice sizes would
be required to avoid the severe finite size effects that would supervene
in simulations performed close to the critical point $\theta=\pi$.

A new fast cluster algorithm was purposely introduced to simulate the
theory with an imaginary $\theta$ term. It works for the operator $Q^{(1)}$.

A salient outcome of our work is the good performance of the
analytic continuation from imaginary to real $\theta$. No theoretical
prejudices were assumed in the functional form used in the extrapolation
and different functions led to comparable results (see Fig.~\ref{Fig2}).

\vskip 10mm
\begin{figure}[htbp]
\centerline{\epsfig{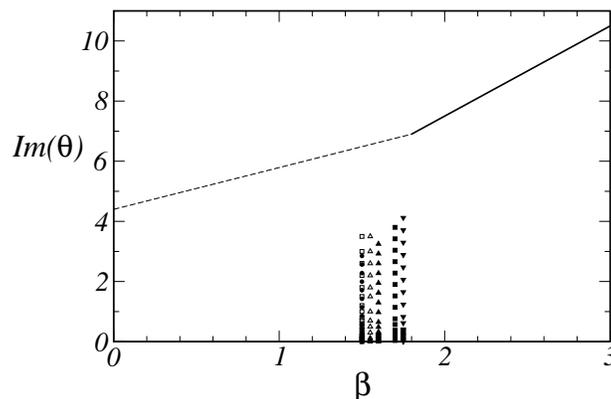}}
\caption{Possible phase diagram in the $Im(\theta)$--$\beta$ plane,
($Im(\theta)\equiv\vartheta$).
The phase transition lines (continuous line is first order and dashed
line is at most second order) are borrowed from Ref.~\protect\cite{bhanot}.
Symbols are situated at the positions where single simulations were
done. Filled (white) symbols mean simulations with $Q^{(1)}$ ($Q^{(2)}$) and
the vertical sequences of data correspond, from left to right,
to $\beta=1.5$, 1.55, 1.6, 1.7 and 1.75. In
this diagram extrapolations appear as going downwards starting from the
Monte Carlo data points, hence they never touch the phase transition lines.}
\label{Fig5}
\end{figure}

A key ingredient
for the successful extrapolation was to have got data from simulations
within a wide range of (imaginary) values of $\overline{\theta}$ for all $\beta$,
($\overline{\vartheta}\equiv-i\overline{\theta}\in\left[0,10\right]$
when $Q^{(1)}$ was
used and $\overline{\vartheta}=\vartheta\equiv -i\theta\in\left[0,3.5\right]$
for $Q^{(2)}$,
the difference of intervals being due to the effect of the nontrivial
renormalization that must be applied to
the former). We also noticed that at very large values of imaginary $\overline{\theta}$ the
extraction of the correlation length becomes more imprecise since the kind of plateaux
shown in Fig.~\ref{Figmass} shrink making it difficult to decide the correct result
for $\xi$ and its error bar. This problem appears approximately at
$\overline{\vartheta}\geq 15$ for the $Q^{(1)}$ charge density.
Then, the method consists in taking data in a wide range of values of imaginary
$\overline{\theta}$ (or $\theta$)
while staying not too far from the real axis in order to make the
extrapolation sensitive to the real physics of the problem and in order to have
wide enough plateaux which allow to easily
extract the correlation length $\xi$.

Moreover our extrapolations did not cross any nonanalytical region
in the phase diagram of the model. Indeed in Fig.~\ref{Fig5}
we show a schematic reproduction of the phase diagram that appears
in Ref.~\cite{bhanot}. The first order transitions lie on the
continuous straight line while the dashed line indicates less
singular transitions, at most second order. Our simulations were
performed at the positions indicated by the symbols
(same symbols than in Figs.~\ref{Fig3} and~\ref{Fig4}; notice that
this plot refers to the values of $\theta$, not of $\overline{\theta}$,
for both topological charge density definitions)
and the corresponding analytic continuations proceed downwards, thus very
far from any possible line of singular points.

{}For $\theta >\pi$ the model should acquire again a finite correlation
length since the Hamiltonian is a periodic function of $\theta$. However
an analytical continuation could hardly display such a behavior beyond
$\theta=\pi$ because of the nonanalyticity at that value of $\theta$.

\section{Acknowledgements}

It is a pleasure to thank Juan Jos\'e Alonso for a critical reading of a draft
of the paper. We are indebted to CINECA (Italy) for the use of their HPC system
for part of our production runs.



\begin{references}
\bibitem{haldane1} F. D. M. Haldane, Phys. Lett. {\bf 93A}, 464 (1983).
\bibitem{haldane2} F. D. M. Haldane, Phys. Rev. Lett. {\bf 50}, 1153 (1983).
\bibitem{lieb} E. H. Lieb, T. Schultz and D. Mattis, Ann. Phys. (N.Y.) {\bf 16}, 407 (1961).
\bibitem{affleck1} I. Affleck and E. H. Lieb, Lett. Math. Phys. {\bf 12}, 57 (1986).
\bibitem{affleck2} I. Affleck, T. Kennedy, E. H. Lieb and H. Tasaki,
Phys. Rev. Lett. {\bf 59}, 799 (1987).
\bibitem{shankar} R. Shankar and N. Read, Nucl. Phys. {\bf B336}, 457 (1990).
\bibitem{zamolodchikov} A.~B. Zamolodchikov and Al.~B.~Zamolodchikov,
Nucl. Phys. {\bf B379}, 602 (1992).
\bibitem{hasenfratz} P. Hasenfratz, M. Maggiore and F. Niedermayer,
Phys. Lett. B {\bf 245}, 522 (1990).
\bibitem{alles1} B. All\'es, A. Buonanno and G. Cella, Nucl. Phys. {\bf B500},
  513 (1997).
\bibitem{fradkin} E. Fradkin, ``Field Theories of Condensed Matter Systems'',
(Addison--Wesley, Reading MA, 1991).
\bibitem{bietenholz} W. Bietenholz, A. Pochinsky and U.--J. Wiese,
  Phys. Rev. Lett. {\bf 75}, 4524 (1995).
\bibitem{azcoiti} V. Azcoiti, G. Di Carlo and A. Galante,
  Phys. Rev. Lett. {\bf 98}, 257203 (2007).
\bibitem{PRDAllesPapa} B. All\'es and A. Papa, Phys. Rev. D {\bf 77}, 056008 (2008).
\bibitem{rajaraman} See for instance R. Rajaraman, ``Solitons and Instantons'',
(North-Holland, Amsterdam, 1982).
\bibitem{papa} A. Di Giacomo, F. Farchioni, A. Papa and E. Vicari, Phys.
  Rev. D {\bf 46}, 4630 (1992).
\bibitem{berg} B. Berg and M. L\"uscher, Nucl. Phys. {\bf B190}, 412 (1981).
\bibitem{pisa1} M. Campostrini, A. Di Giacomo and H. Panagopoulos,
  Phys. Lett. B {\bf 212}, 206 (1988).
\bibitem{vicari} B. All\'es and E. Vicari, Phys. Lett. B {\bf 268}, 241 (1991).
\bibitem{pisa2} A. Di Giacomo and E. Vicari, Phys. Lett. B {\bf 275}, 429 (1992);
  B. All\'es, M. Campostrini, A. Di Giacomo, Y. G\"und\"u\c{c} and
  E. Vicari, Phys. Rev. D {\bf 48}, 2284 (1993).
\bibitem{farchioni}  F. Farchioni and A. Papa, Nucl. Phys. {\bf B431}, 686 (1994).
\bibitem{beccalles}  B. All\'es, M. Beccaria and F. Farchioni,
 Phys. Rev. D {\bf 54}, 1044 (1996).
\bibitem{belavin} A. A. Belavin and A. M. Polyakov, JETP Lett. {\bf 22}, 245 (1975).
\bibitem{rots} A rotation of $\pi/4$ radians in $x$--space; another rotation of
$\pi/2$ radians around the $1$--axis in O(3) group space and a parity inversion in the
O(3) space again on the resultant $3$--axis (this last inversion allows to end up
with a topological charge $Q=+1$ object).
\bibitem{creutz} M. Creutz, Phys. Rev. D {\bf 21}, 2308 (1980).
\bibitem{spencer} A. Jevicki, Nucl. Phys. {\bf B127}, 125 (1977);
C. Michael and P.S. Spencer, Phys. Rev. D {\bf 50}, 7570 (1994).
\bibitem{teper} M. Teper, Phys. Lett. B {\bf 171}, 81 (1986); B {\bf 171}, 86 (1986).
\bibitem{cosmai} See B. All\'es, L. Cosmai, M. D'Elia and A. Papa,
Phys. Rev. D {\bf 62}, 094507 (2000) and references therein.
\bibitem{luscher} M. L\"uscher, Commun. Math. Phys. {\bf 85}, 39 (1982).
\bibitem{wolff} U. Wolff, Phys. Rev. Lett. {\bf 62}, 361 (1989).
\bibitem{swendsen} R. Swendsen and J.--S. Wang, Phys. Rev. Lett. {\bf 58}, 86 (1987).
\bibitem{lauwers} P. G. Lauwers and V. Rittenberg, Phys. Lett. B {\bf 233}, 197 (1989).
\bibitem{wang} J.--S. Wang, Physica A (Amsterdam) {\bf 161}, 249 (1989).
\bibitem{dimi} I. Dimitrovic, P. Hasenfratz, J. Nager and F. Niedermayer,
  Nucl. Phys. {\bf B350}, 893 (1991).
\bibitem{metropolis} N. Metropolis, A. Rosenbluth, M. Rosenbluth, A. Teller and E. Teller,
J. Chem. Phys. {\bf 21}, 1087 (1953).
\bibitem{petronzio} R. Petronzio and E. Vicari, Phys. Lett. B {\bf 254}, 444 (1991).
\bibitem{fortuin} C. M. Fortuin and P. W. Kasteleyn, Physica (Amsterdam) {\bf 57}, 536 (1972).
\bibitem{hoshen} J. Hoshen and R. Kopelman, Phys. Rev. B {\bf 14}, 3438 (1976).
\bibitem{niedermayer} F. Niedermayer, Phys. Lett. B {\bf 237}, 473 (1990).
\bibitem{controzzi} D. Controzzi and G. Mussardo, Phys. Rev. Lett. {\bf 92},
021601 (2004); Phys. Lett. B {\bf 617}, 133 (2005); L. Campos Venuti,
C. Degli Esposti Boschi, E. Ercolessi, F. Ortolani, G. Morandi, S. Pasini and M. Roncaglia,
J. Stat. Mech. (2005) L02004.
\bibitem{kronfeld} A. S. Kronfeld, Nucl. Phys. Proc. Suppl. {\bf 17}, 313 (1990);
  M. L\"uscher and U. Wolff, Nucl. Phys. {\bf B339}, 222 (1990).
\bibitem{wolff2} U. Wolff, Nucl. Phys. Proc. Suppl. {\bf 17}, 93 (1990).
\bibitem{improestimatorscomment} Had we decided to make
use of improved estimators (at least for the two--point correlation functions),
the only algorithm that would have allowed us to use them is the
Wang algorithm~\cite{wang}; if instead we
had decided on the Lauwers--Rittenberg method~\cite{lauwers}, then the calculation
of improved estimators would have become again intractable even for simple
two--point correlators $\langle\vec{\phi}(x)\cdot\vec{\phi}(y)\rangle$.
In fact, in such a case the expression of the estimator when $x,y$
belong to different clusters (say $C_x$ and $C_y$)
does not vanish but comes out proportional
to $\tanh(\overline{h}_x)\,\tanh(\overline{h}_y)$ with
$\overline{h}_x\equiv\sum_{z\in C_x} h(z) \vert\vec{\phi}(z)\cdot \vec{r}\vert$
(and analogously for $\overline{h}_y$)
where $h(z)$ is the local magnetic field, Eq.(\ref{fieldmag}).
Clearly such an estimator does not improve the calculation since it can take
any sign and moreover it requires a double sum on clusters which makes the
necessary computer time exceedingly large.
\bibitem{affleck3} I. Affleck and F. D. M. Haldane, Phys. Rev. B {\bf 36}, 5291 (1987);
I. Affleck, D. Gepner, H. J. Schulz and T. Ziman, J. Phys. A {\bf 22}, 511 (1989),
(Erratum: ibid. A {\bf 23}, 4725 (1990)).
\bibitem{bhanot} G. Bhanot and F. David, Nucl. Phys. {\bf B251}, 127 (1985).
\end{references}
\end{document}